\documentclass[prx, letterpaper, superscriptaddress, amsmath, amssymb, amssymb, reprint, floatfix, nofootinbib]{revtex4-2}

\usepackage{graphicx}
\usepackage{dcolumn}
\usepackage{bm}
\usepackage{xcolor} 
\usepackage[normalem]{ulem}
\usepackage{braket}

\newcommand{\parens}[1]{\left({#1}\right)}

\begin{document}

\preprint{APS/123-QED}


\title{Floquet interferometry of a dressed semiconductor quantum dot}

\author{Felix-Ekkehard von Horstig}
\altaffiliation{These authors contributed equally to this work}
 \affiliation{Quantum Motion, 9 Sterling Way, London, N7 9HJ, United Kingdom}
 \affiliation{Department of Materials Sciences and Metallurgy, University of Cambridge, Charles Babbage Rd, Cambridge CB3 0FS, United Kingdom}
 
\author{Lorenzo Peri${}^\dagger$}
\altaffiliation{These authors contributed equally to this work}
\affiliation{Quantum Motion, 9 Sterling Way, London, N7 9HJ, United Kingdom}
\affiliation{Cavendish Laboratory, University of Cambridge, JJ Thomson Ave, Cambridge CB3 0HE, United Kingdom}
 
 \author{Sylvain Barraud}
 \affiliation{CEA, LETI, Minatec Campus, F-38054 Grenoble, France}

  \author{Sergey~N.~Shevchenko}
 \affiliation{B. Verkin Institute for Low Temperature Physics and Engineering, Kharkiv 61103, Ukraine}
 
 \author{Christopher J. B. Ford}
 \affiliation{Cavendish Laboratory, University of Cambridge, JJ Thomson Ave, Cambridge CB3 0HE, United Kingdom}
 
 \author{M. Fernando Gonzalez-Zalba${}^{\ddagger}$}
 \affiliation{Quantum Motion, 9 Sterling Way, London, N7 9HJ, United Kingdom}
 \email{fernando@quantummotion.tech}

\date{\today}

\begin{abstract}

A quantum system interacting with a time-periodic excitation creates a ladder of hybrid eigenstates in which the system is mixed with an increasing number of photons. This mechanism, referred to as dressing, has been observed in the context of light-matter interaction in systems as varied as atoms, molecules and solid-state qubits. In this work, we demonstrate state dressing in a semiconductor quantum dot tunnel-coupled to a charge reservoir. We observe the emergence of a Floquet ladder of states in the system's high-frequency electrical response, manifesting as interference fringes at the multiphoton resonances despite the system lacking an avoided crossing. We study the dressed quantum dot while changing reservoir temperature, charge lifetime, and excitation amplitude and reveal the fundamental nature of the mechanism by developing a theory based on the quantum dynamics of the Floquet ladder, which is in excellent agreement with the data. Furthermore, we show how the technique finds applications in the accurate electrostatic characterisation of semiconductor quantum dots.

\end{abstract}

\maketitle
\def\thefootnote{$\dagger$}\footnotetext{lp586@cam.ac.uk}\def\thefootnote{\arabic{footnote}}
\def\thefootnote{$\ddagger$}\footnotetext{fernando@quantummotion.tech}\def\thefootnote{\arabic{footnote}}

\section{Introduction}

Under periodic driving of a quantum system, its state becomes \textit{dressed} by the drive tone (signal). This can be pictured as the system being described by an infinite \textit{ladder} of states, in which the system is hybridized with an increasing number of photons~\cite{tsujiFloquetStates2024,Rudner_Lindner_2020,Rodriguez_2018}. 
Coherent dressing of quantum systems has been investigated in a multitude of platforms, such as electron~\cite{lauchtDressedSpin2017,Koski2018}, hole~\cite{boscoPhaseDriving2023} and nuclear~\cite{londonDetectingPolarizing2013} spins, self-assembled quantum dots (QDs)~\cite{xuCoherentOptical2007} and superconducting qubits~\cite{Wilson2007,Berns2008,baurMeasurementAutlerTownes2009}, and is an enticing candidate for quantum coherent control~\cite{Oka_2019,Rodriguez-Vega_2021,Castro_2023,Shevchenko_2014_ampl}.

Thus far, however, most theoretical and experimental efforts have focused on coupled two-level systems, periodically driven across an avoided crossing. This leads to a finite occupation of the excited state via diabatic Landau-Zener transitions and the accumulation of a dynamical phase, giving rise to Landau-Zener-St\"uckelberg-Majorana (LZSM) interference~\cite{Shevchenko2010,Ivakhnenko_2023}. 

In this work, in contrast to observations of LZSM interference in coupled QD systems~\cite{Petta2010, Stehlik2012, Dupont-Ferrier2013, Anasua2018}, we study the effect of dressing a single charge level housed in a QD and coupled to a reservoir. We probe the energy spectrum of the dressed state by measuring its dynamical response to a probe tone, resonant with a superconducting cavity coupled to the QD. The result is the appearance of interference fringes in the electrical response of the system, which we term Floquet interferometry. This reveals the presence of coherent interference despite the fact that charges tunnelling to the reservoir randomize the charge phase. We find that the interference fringes depend on dressing amplitude and frequency, as well as the relation between the width of the undressed QD response and the photon energy of the dressing frequency. Unlike in other similar experiments, our system does not contain any avoided crossing~\cite{peyruchat2024landauzener,Wen_2020,Gong_2016,Chen_2021,Liul_2023}, and the interference is thus solely due to the Floquet dynamics of a single level. This work directly probes its underlying Floquet state structure, made visible by stochastic transitions of charges between the QD and reservoir.  

Overall, we demonstrate the Floquet interferometry signal of the (singly) dressed QD in the small-probe regime for varying (i) charge reservoir hole temperature and (ii) lifetime of the charge state in the QD. We further explore the effect of doubly dressing the QD with a strong probe tone resonant with the superconducting cavity. The excellent agreement with the Floquet theory elucidates the underlying structure of the Floquet ladders, casting light on their respective interaction and their effect on the stochastic QD--reservoir charge tunnelling. Furthermore, we reveal how this technique can be utilised for a simple, accurate and complete characterisation of the QD's electrostatic properties. 

\section{Coherent interaction of a dressed Quantum Dot with a reservoir}

\begin{figure*}
    \centering
    \includegraphics[width=1\textwidth]{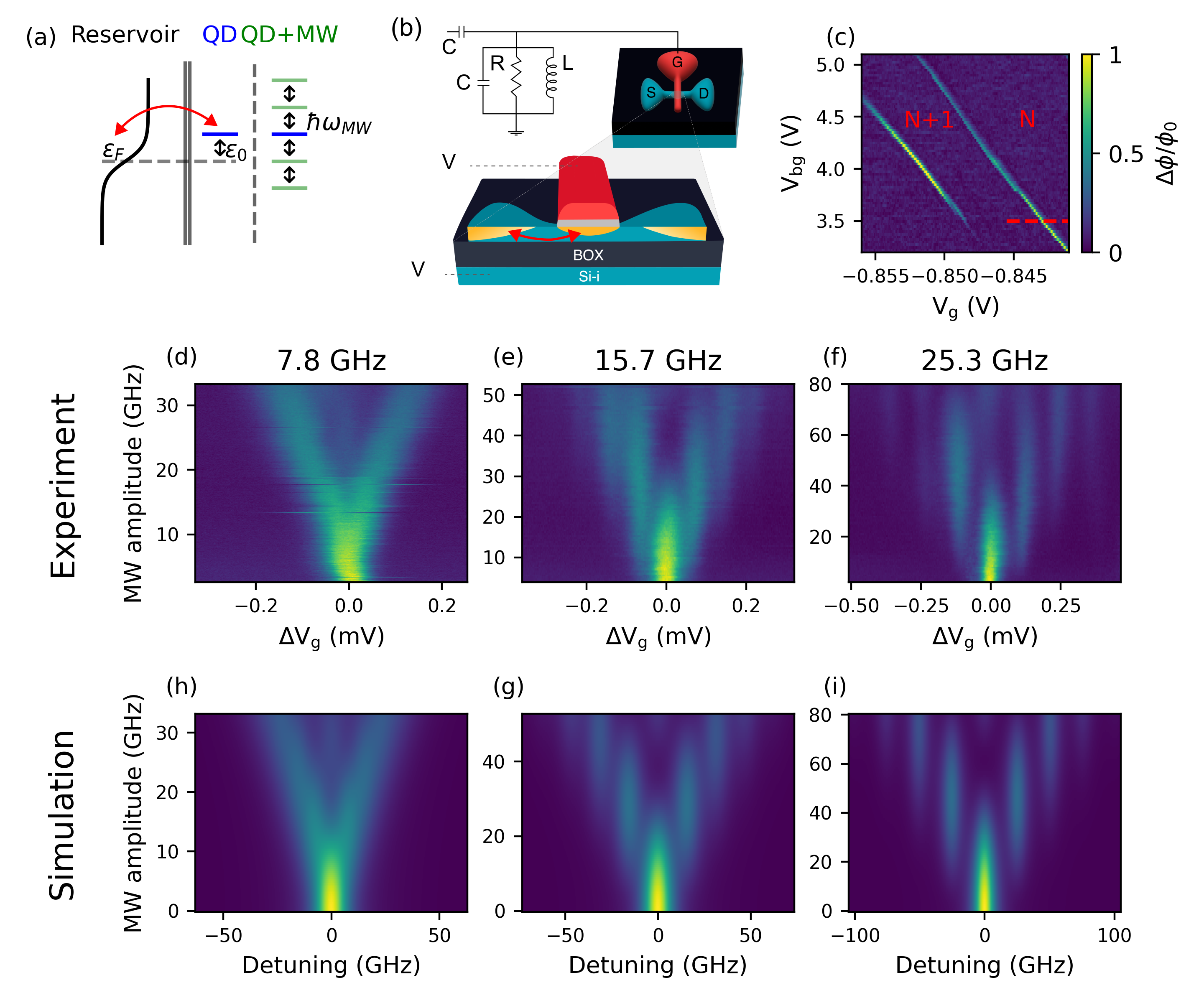}
    \caption{(a) Energy-level schematic showing a reservoir and a QD. When a microwave tone is applied to the QD it splits into a ladder of Floquet states separated by the photon energy $\hbar\omega_\text{MW}$. (b) Device schematic showing a single-gate silicon nanowire-on-insulator transistor embedded in a lumped-element resonant circuit. The cross-section shows the QD formed under the gate and the Source and Drain reservoirs. (c) Stability diagram showing two dot--reservoir transitions. We focus on the transition on the right between (nominal) charge occupations of $N$ and $N+1$ and measure along the dashed line indicated. (d-f) Experimental Floquet interferometry measurements carried out at three different frequencies, as indicated. (h-i) Simulated response of the interferometry matching the corresponding data above.}
    \label{fig:LZ-vs-freq}
\end{figure*}

As is well known, Floquet's theorem states that the eigenstates of a periodically driven quantum system will themselves be periodic in time~\cite{Rudner_Lindner_2020,Koski2018,Tucker_Feldman_1985}. In particular, Tien and Gordon showed how the Floquet eigenstate of a single level (the QD) driven sinusoidally as $E(t) = \varepsilon_0 + \delta \varepsilon \cos{\omega t}$ can be written in the form~\cite{Tien_Gordon_1963}
\begin{equation}
    \ket{\phi(t)} = e^{{\rm i}\varepsilon_0t/\hbar} \sum_{m=-\infty}^{\infty} J_m\left( \frac{\delta\varepsilon}{\hbar\omega}\right)
    e^{{\rm i}m\omega t} \ket{\phi_m} ,
    \label{eq:Floquet_State}
\end{equation}
\noindent 
where $\varepsilon_0$ represents the (on-site) energy of the QD, and $J_m$ is the $m$-th Bessel function of the first kind. $\ket{\phi_m}$ represents the Fourier decomposition of the Floquet state, where we recognize the Jacobi-Anger decomposition of the oscillating phase of the dressed single level, i.e., $\exp\left({\rm i}\delta\varepsilon\sin{\omega t}/\hbar \omega\right)$.
Crucially, we can see how Eq.~(\ref{eq:Floquet_State}) effectively forms a ladder of equally spaced states with (quasi-)energies given by $\varepsilon_0 + m\hbar \omega$, and for brevity we shall therefore refer to the Fourier coefficients $\ket{\phi_m}$ as Floquet \textit{rungs} (Fig.~\ref{fig:LZ-vs-freq}a). Informally, the reader may picture the sinusoidal drive that \textit{dresses} the single state as creating \textit{copies} $\ket{\phi_m}$ of the original state $\ket{\phi}$. A charged particle in the dressed QD will then spread into a coherent superposition of the rungs, which will be occupied with probability $\braket{\phi_m|\phi_m}=J_m^2\left( \delta\varepsilon/\hbar\omega\right)$ and whose phase will oscillate in time according to its quasi-energy. Notably, this phase directly depends on the photon number $m$, and, therefore, different rungs will be able to interfere coherently with each other.

In order to probe this interaction, we can couple the dressed QD with a charge reservoir and dispersively measure its interaction with a radio-frequency (RF) cavity with resonant frequency $\omega_\text{RF}$ (Fig~\ref{fig:LZ-vs-freq}b). For a p-doped reservoir, this configuration is commonly referred to in the literature as a single-hole box (SHB)~\cite{Oakes2023}. In Ref.~\cite{Peri_2024}, we show how such a system can be effectively modelled as a driven two-level system (whose states are the occupied and unoccupied QD), obeying the time-dependent Hamiltonian
\begin{equation}
    \begin{aligned}
    &H(t) = H_0 + H_\text{RF} + H_\text{MW} = \\
    &=\frac{\varepsilon_0}{2} \sigma_z + \frac{\delta \varepsilon_\text{RF}}{2} \sigma_z \cos{\omega_\text{RF} t} + \frac{\delta \varepsilon_\text{MW}}{2} \sigma_z \cos{\omega_\text{MW} t} ,
    \end{aligned}
    \label{eq:Hamiltonian}
\end{equation}
\noindent
where we label the driving tone as MW to distinguish it from the RF excitation used to probe the system via the resonator. The (incoherent) interaction with the reservoir is then modelled via a Lindblad master equation, where we consider the possibility of stochastically tunnelling into (or out of) the QD with rate $\Gamma_+(t)$ ($\Gamma_-(t)$), which depend on time via the drive and the probe tones. The (driven) charge tunnelling events will create an AC gate current which will be picked up by the resonator, allowing us to gain information on the state of the system.

In the main text, we will focus on the physical explanation of the observed phenomenon, striving to provide an intuitive understanding. The reader interested in the mathematical details, or who would enjoy a refresher on Floquet theory, is welcome to browse Appendix~\ref{app:Maths}, which contains the extension of Ref.~\cite{Peri_2024} to the MW-dressed case.

The incarnation of the system described above is a QD formed under a silicon-on-insulator nanowire transistor, coupled (with tunnel rate $\Gamma$) to a p-type reservoir (with hole temperature $T_\text{h}$) (Fig.~\ref{fig:LZ-vs-freq}b). By monitoring the reflected signal of a superconducting LC resonator of frequency $\omega_\text{RF}/2\pi = 2.1$\,GHz connected to the top gate, we detect the gate current arising from single-hole transitions between the QD and the reservoir. In particular, the dispersive signal will be proportional to the effective admittance $Y$ of the SHB at the resonator frequency~\cite{Peri_2024,peri2023unified,oakes2023multiplier}.
By applying voltages on the top gate (V$_\text{g}$) and back-gate (V$_\text{bg}$) separately, we find several dot--reservoir transitions (DRTs). We focus on one such transition, as indicated in Fig.~\ref{fig:LZ-vs-freq}c.

To begin with, we consider the case of a small RF probe signal ($\delta \varepsilon_\text{RF} \ll k_\text{B} T_\text{h}, h \Gamma$).
We apply the microwave drive of frequency $\omega_\text{MW}$ to the source of the device (here employed as an additional gate) and monitor the phase response of the resonator as a function of gate voltage as we increase the power of the MW tone.
At low frequencies ($\omega_\text{MW}/2\pi = 7.8$\,GHz, Fig.~\ref{fig:LZ-vs-freq}d), we mostly observe a broadening of the transition. However, as the frequency is increased to 15.7\,GHz (Fig.~\ref{fig:LZ-vs-freq}e) and 25.3\,GHz (Fig.~\ref{fig:LZ-vs-freq}f), interference fringes become increasingly more apparent. 
These fringes arise from the phase interactions of the different rungs of the Floquet ladder, as discussed above. 

More quantitatively, the small-signal admittance of an (undressed) SHB takes the form of a single peak,~\cite{Peri_2024}
\begin{equation}
    Y_0(\varepsilon_0) = \frac{(\alpha e)^2}{2 k_\text{B} T_\text{h}} \frac{\Gamma \omega_\text{RF}}{\omega_\text{RF} - \text{i} \Gamma} \mathcal{F}'(\varepsilon_0) ,
    \label{eq:Y0_undressed}
\end{equation}
\noindent where, assuming (as is the case in this work) that $\hbar \omega_\text{RF} \ll k_\text{B} T_\text{h}, h \Gamma$,
\begin{equation}
    \mathcal{F}'(\varepsilon_0)= \cosh^{-2} {\left(\frac{\xi}{2 k_\text{B} T_\text{h}}\right)} * \frac{\Gamma/\pi}{\Gamma^2 + (\xi - \varepsilon_0)^2 }
    \label{eq:F_broad}
\end{equation}
\noindent is the convolution between the (derivative of the) Fermi-Dirac distribution in the reservoir and the effective (Lorentzian) density of states of the QD whose state becomes metastable (with lifetime $\Gamma$) because of the coupling with the reservoir. The subscript 0 in Eq.~(\ref{eq:Y0_undressed}) is used to highlight the fact that it represents the admittance of the system when not subject to the MW drive ($H_\text{MW}=0$), but only to a small RF probe.

We can now recall the ansatz developed above of the MW dressing creating \textit{copies} of the original system, equally spaced in energy space by a photon energy. For increasing MW amplitude $\delta \varepsilon_\text{MW}$, further rungs will subsequently be populated according to Eq.~(\ref{eq:Floquet_State}). Each of them will then be perturbed by the (small) probe, which will cause a gate current at frequency $\omega_\text{RF}$ for each $\ket{\phi_m}$, without any mixing of the different rungs, at least to first order in $\delta \varepsilon_\text{RF}$. It thus becomes tempting to write
\begin{equation}
    Y(\varepsilon_0) = \sum_{m=-\infty}^{\infty} J_m^2 \left( \frac{\delta\varepsilon_\text{MW}}{\hbar\omega_\text{MW}}\right) Y_0(\varepsilon_0 + m \hbar \omega_\text{MW}),
    \label{eq:Y_MW}
\end{equation}
\noindent
which simply represents the DRT peak corresponding to a single rung $\ket{\phi_m}$ coupled to a reservoir (centered at its quasi-energy) multiplied by how `likely' the rung is to be occupied. This is the same result we arrive at after the precise calculations in Appendix~\ref{app:Maths}. 


Comparing Eq.~(\ref{eq:Y_MW}) with the experimental data (Fig.~\ref{fig:LZ-vs-freq}h-i), we find excellent agreement at all three frequencies. However, with the insight provided by the model, we can explain the qualitative difference between the low- and high-frequency cases. When the MW photon energy is smaller than the width of the peaks, it is not possible to resolve the separate rungs, resulting in a broadening for intermediate power, and interference fringes as higher quasi-energies become populated. The opposite occurs for large MW frequency, where it becomes possible to distinguish the single-photon lines. This will be explored quantitatively in the following.

Notably, it is thus possible to utilize the separation of the peaks at high frequency to find the lever arm of the top-gate on the QD, defined as $\alpha=C_\text{g}/C_\Sigma$, where $C_\text{g}$ is the gate capacitance and $C_\Sigma$ is the total capacitance of the QD, which also contains the capacitances to the charge reservoirs and the back-gate. We find $\alpha=0.89\pm 0.02$, where we have used the equation $e\alpha\Delta V_\text{g} = \Delta\varepsilon = N\hbar \omega_\text{MW}$, $e$ and $\Delta V_\text{g}$ being the electron charge and the voltage separation of the photon lines, respectively. The use of the technique for electrostatic characterisation of the QD will also be explored further in the following sections.

To further characterize the interactions of the QD with the charge reservoir, we investigate the dependence of broadening on: 1) increasing tunnel rates, which causes lifetime broadening of the QD energy level, 2) increasing temperature, which produces thermal broadening due to the reservoir, and 3) increasing RF measurement power, which produces broadening caused by a finite dressing of the QD at $\omega_\text{RF}$ as well as at $\omega_\text{MW}$.

\subsection{Tunnel-rate Dependence}
\begin{figure*}
    \centering
    \includegraphics[width=1\textwidth]{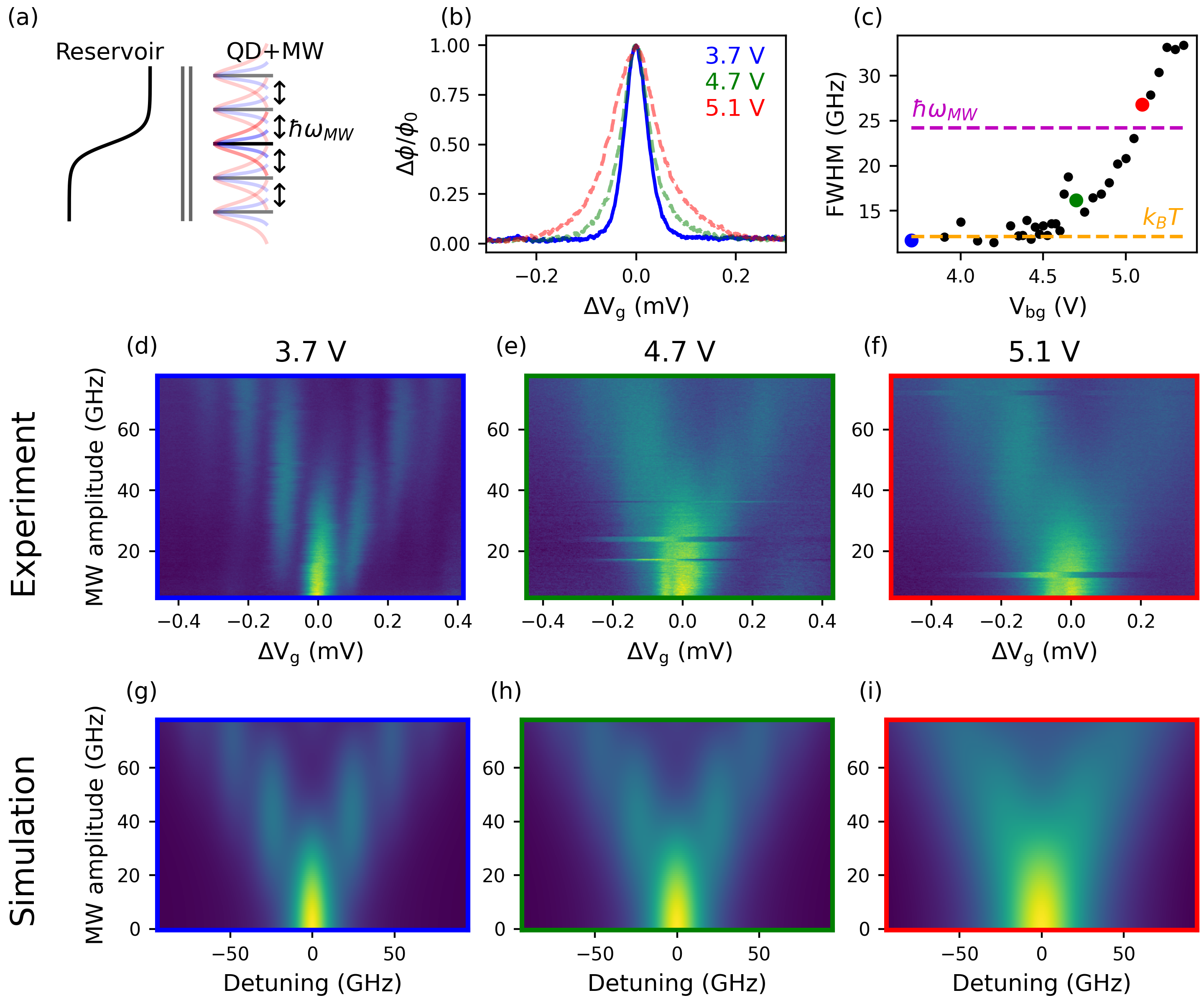}
    \caption{Effect of lifetime broadening on Floquet interferometry: (a) energy-level schematic showing the reservoir and a Floquet ladder of QD + MW states. Lifetime broadening produces a broadening of the Floquet states. (b) Phase response of the DRT measured at three different V$_\text{bg}$, as indicated. (c) FWHM extracted as a function of V$_\text{bg}$. The three measurement points (3.7\,V, 4.7\,V and 5.1\,V) are indicated in blue, green and red, respectively. For reference, we also indicate the thermal energy $k_\text{B}T$ and the photon energy $\hbar\omega_\text{MW}$. (d-f) Experimental measurements of Floquet interferometry carried out with $\omega_\text{MW}/2\pi = 24.2$\,GHz at three different V$_\text{bg}$. (g-h) Simulated response corresponding to the measurements above.}
    \label{fig:LZ-vs-Bg}
\end{figure*}

We utilize the back-gate of the transistor to tune the tunnel rate of the DRT. An increasing tunnel rate produces a change in the small-signal admittance $Y_0$. In particular, when the broadening of the (metastable) QD state due to its finite lifetime is negligible with respect to the thermal broadening of the reservoir ($h \Gamma \ll k_\text{B} T_\text{h}$), its full-With half-maximum (FWHM) reads $\textnormal{FWHM}_0 = 3.53 k_\text{B} T_\text{h}$~\cite{cottet2011,Ahmed2018,Peri_2024}. As the tunnel rate increases, the Lorentzian becomes broader (see Fig.~\ref{fig:LZ-vs-Bg}a) and dominates in Eq.~(\ref{eq:F_broad}). Thus~\cite{Ahmed2018b,Peri_2024}
\begin{equation}
    \textnormal{FWHM}_0 \approx \sqrt{(3.53 k_\text{B} T_\text{h})^2 + (h \Gamma)^2} ,
\end{equation}
\noindent
leading to lifetime broadening when $\Gamma \gtrsim k_\text{B} T_\text{h}$.

We first characterize the effect of V$_\text{bg}$ on the FWHM of the DRT by fitting the resonator response of voltage traces recorded at different V$_\text{bg}$ (see Fig.~\ref{fig:LZ-vs-Bg}b). We plot the FWHM against V$_\text{bg}$ in Fig.~\ref{fig:LZ-vs-Bg}c and find that at low V$_\text{bg}$ the FWHM is thermally limited, finding $T_\text{h} \sim$ 160 mK as the hole temperature. Above V$_\text{bg} \sim$ 4.5~V the DRT becomes lifetime broadened, indicating that we are able to tune the FWHM by a factor of approximately 4.

We now apply a microwave tone of frequency $\omega_\text{MW}/2\pi = 24.2$\,GHz, and again monitor the phase response of the resonator as a function of gate voltage and MW power. At V$_\text{bg} = 3.7$\,V, a clear interference pattern can be observed (Fig.~\ref{fig:LZ-vs-Bg}c). As $\Gamma$ is increased, the interference pattern is increasingly washed out by the broadening of the DRT (Fig.~\ref{fig:LZ-vs-Bg}d-e), in agreement with the simulations (Fig.~\ref{fig:LZ-vs-Bg}g-i).

The physical explanation of this trend is easily found by considering Fig.~\ref{fig:LZ-vs-Bg}a. Recalling the earlier discussion, when the photon energy is larger than the width of $Y_0$ ($\hbar \omega_\text{MW} \gg \sqrt{(3.53 k_\text{B} T_\text{h})^2 + (h \Gamma)^2}$), the photon lines can be clearly resolved (Fig.~\ref{fig:LZ-vs-Bg}d,g). As we increase $\Gamma$, however, the rungs of the Floquet ladder become more washed out, blurring the peaks into each other. As FWHM${}_0$ approaches and surpasses $\hbar \omega_\text{MW}$ (see Fig.~\ref{fig:LZ-vs-Bg}c), the visibility of the lines is completely lost (Fig.~\ref{fig:LZ-vs-Bg}f,i).

\subsection{Temperature Dependence}
\begin{figure*}
    \centering
    \includegraphics[width=1\textwidth]{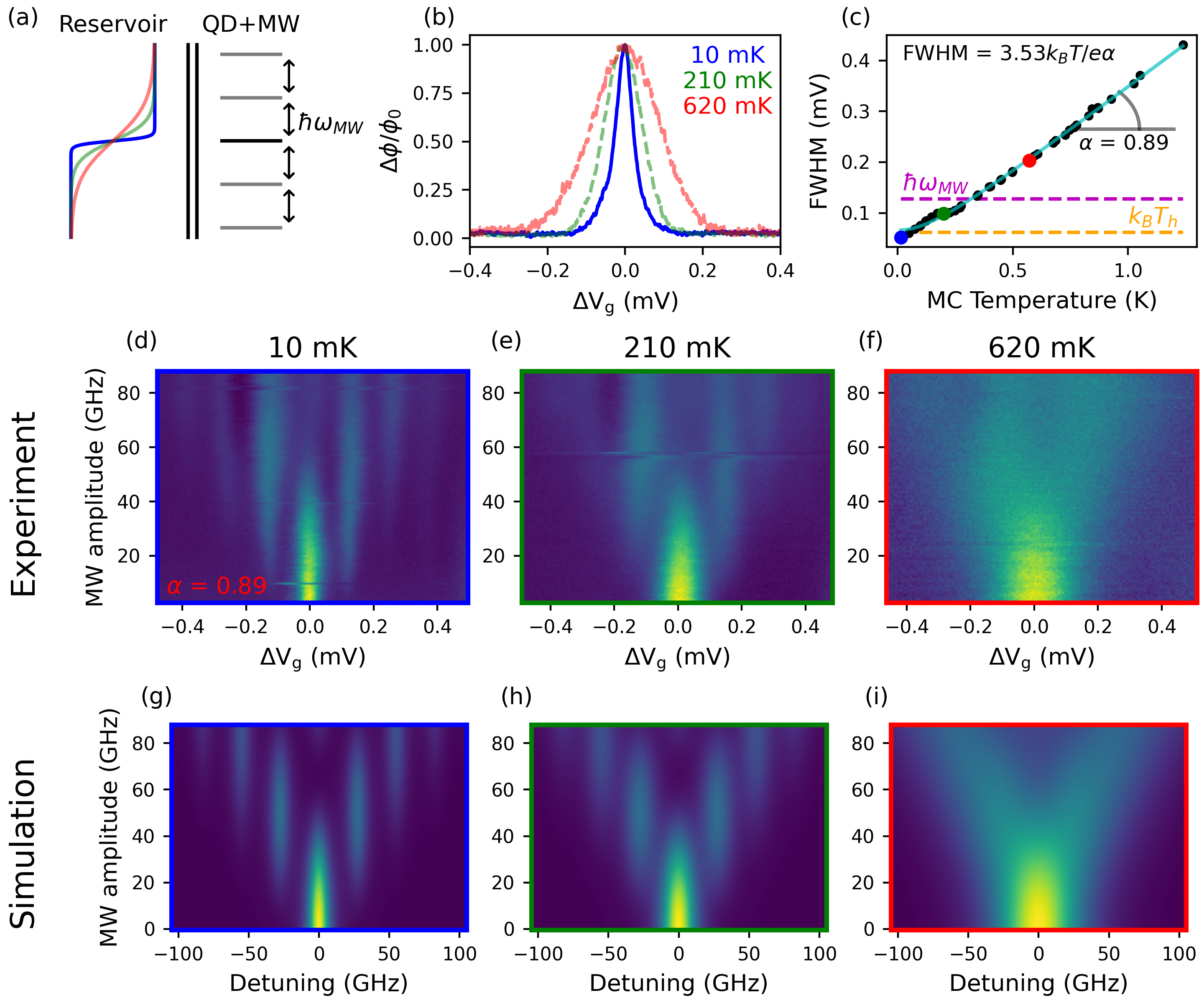}
    \caption{Effect of thermal broadening on Floquet interferometry: (a) energy-level schematic showing the reservoir and a Floquet ladder of QD + MW states. Thermal broadening produces a broadening of the reservoir Fermi surface. (b) Phase response of the DRT measured at three different mixing-chamber temperatures, as indicated. (c) FWHM extracted as a function of mixing-chamber temperature $T_{MC}$. The three measurement points (10\,mK, 210\,mK and 620\,mK) are indicated in blue, green and red, respectively. For reference, we also indicate the hole temperature $k_\text{B}T_\text{h}$ and the photon energy $\hbar\omega_\text{MW}$. (d-f) Experimental measurements of Floquet interferometry carried out with $\omega_\text{MW}/2\pi = 27.5$\,GHz at the three different temperatures. (g-h) Simulated response corresponding to the above measurements.}
    \label{fig:LZ-vs-Temp}
\end{figure*}

Increasing temperature causes a thermal smearing of the Fermi-Dirac distribution in the reservoir, which results in a broadening of the DRT (Fig.~\ref{fig:LZ-vs-Temp}a)~\cite{Ahmed2018b,oakes2023multiplier,Oakes2023}.

We first characterise this effect by fitting the phase response of the DRT on the resonator at different mixing-chamber temperatures $T_\text{MC}$, as plotted in Fig.~\ref{fig:LZ-vs-Temp}b. As temperature is increased, the FWHM follows~\cite{Ahmed2018b}
\begin{equation}\label{eq:FWHM_T-Th}
    \text{FWHM}_0 \left[V\right] = \frac{3.53 k_\text{B}}{e \alpha} \sqrt{T_\text{MC}^2+T_\text{h}^2} .
\end{equation}
At low $T_\text{MC}$, the FWHM is determined by the hole temperature $T_\text{h}$, the residual thermal energy in the reservoir, while when $T_\text{MC} \gtrsim T_\text{h}$, the FWHM increases linearly (Fig.~\ref{fig:LZ-vs-Temp}c). Notably, when measured in \textit{Volts}, the FWHM has a slope determined by the lever arm $\alpha$~\cite{Ahmed2018b}. 
We thus use Eq.~(\ref{eq:FWHM_T-Th}) to fit the FWHM, and find $T_\text{h} = 160 \pm 40$\,mK and $\alpha = 0.89 \pm 0.01$, in good agreement with the values of $\alpha$ extracted from the separation of the photon lines, and hence validating the technique for electrostatic parameter extraction. 
We note a deviation from the fit at low $T$, which we attribute to charge noise arising from one or more two-level fluctuators, resulting in an asymmetric line shape at low $T_\text{MC}$. This extra noise makes accurate fitting of the FWHM difficult for those datasets. 

Finally, we apply a MW tone of frequency $\omega_\text{MW}/2\pi = 27.5$\,GHz and measure the resonator phase response as a function of gate voltage and MW power at three different temperatures: 10\,mK, 210\,mK and 620\,mK (Fig.~\ref{fig:LZ-vs-Temp}d-f). We observe interference fringes that increasingly smear out as the temperature increases, in good agreement with the theory (Fig.~\ref{fig:LZ-vs-Temp}g-i). The physical explanation is the same as before, with the only distinction being that it is now the distribution in the reservoir, and no longer the effective density of states of the QD, that is being smeared out (Fig.~\ref{fig:LZ-vs-Temp}a).
 
\subsection{Finite RF Probe Amplitude}
\label{sec:large-RF}

\begin{figure*}
    \centering
    \includegraphics[width=1\textwidth]{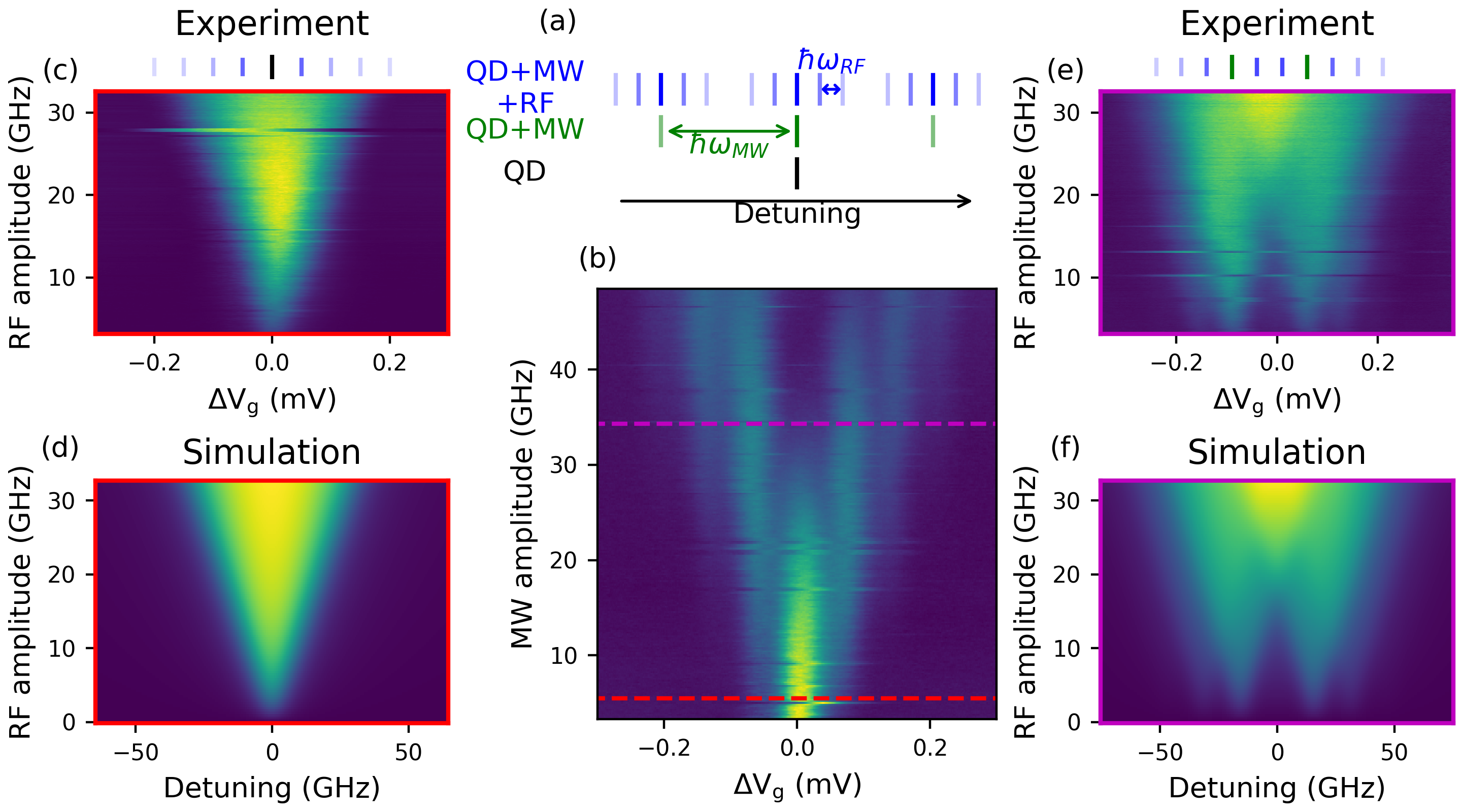}
    \caption{Effect of power broadening on Floquet interferometry. (a) Under sufficient MW and RF power, both the MW and RF tones of the resonator produce a ladder of Floquet states separated in detuning from the energies of the bare QD by the photon energies $\hbar\omega_\text{RF}$ and $\hbar\omega_\text{MW}$. (d) Floquet interferometry carried out at $\omega_\text{MW}/2\pi = 15.7$\,GHz. The measurement in (c) is carried out at zero MW power (red), while (e) is carried out at the MW power indicated (magenta). Experimental data (c) and Simulation (d) of a DRT under increasing RF power. As RF power is increased, power broadening is observed, due to a summation of the Floquet ladder produced by the RF tone of the readout resonator (see dashed lines above (c)). Experimental data (e) and simulation (f) of the DRT under a constant MW tone and increasing RF power. Note that each of the Floquet mode signals is power-broadened and at large power these are summed together when they overlap. }
    \label{fig:Lz-power-broadening}
\end{figure*}

Lastly, we explore the case where the RF probe is no longer small compared to the other broadening parameters ($k_\text{B} T_\text{h}, h \Gamma$). This regime is particularly interesting from a theoretical perspective, as, if $\omega_\text{RF}$ and $\omega_\text{MW}$ are incommensurate, the Hamiltonian of the system is no longer \textit{periodic} in time, so standard Floquet theory does not apply. 
Moreover, we can see how in Eq.~(\ref{eq:Hamiltonian}) $H_\text{RF}$ is exactly identical to $H_\text{MW}$, and so we expect the RF and MW dressing of the QD to have similar effects on the measurement. In Fig.~\ref{fig:Lz-power-broadening}, however, we see that this is clearly not the case. Here, we monitor the resonator response for increasing RF amplitude at two different values of $\delta \varepsilon_\text{MW}$ (Fig.~\ref{fig:Lz-power-broadening}c,e), observing a starkly different effect of the MW drive and the RF probe. The latter, in fact, rather than creating peaks at the different photon lines, separated by interference fringes, results in a \textit{single} peak, which broadens linearly with $\delta \varepsilon_\text{RF}$, a phenomenon known as power broadening~\cite{Peri_2024,oakes2023multiplier,Oakes2023}. 
To understand the physical origin of this different response we can borrow a result from Appendix~\ref{app:Maths} and study the gate current produced by the (doubly dressed) QD exchanging particles with the reservoir, which can be written in the form
\begin{equation}
    I_\text{g}(t) = \alpha e \sum_{N=0}^{\infty} \sum_{M=0}^{\infty} \beta_{M,N} \sin{\left(N\omega_\text{RF}t + M \omega_\text{MW}t + \phi_{M,N}\right)} ,
    \label{eq:gate_current}
\end{equation}
\noindent
where $\tan{\phi_{M,N} = (N\omega_\text{RF} + M \omega_\text{MW})/2 \pi \Gamma}$ and $\beta_{M,N}$ is a superposition of Lorentzian peaks centred at the various combinations of photon energies (weighted by Bessel functions of the first kind) convolved with the Fermi-Dirac distribution of the charge reservoir.
Equation~(\ref{eq:gate_current}) reveals the true character of the current experiment, which can be understood electrically as a quantum \textit{mixer}, where the strong driving of the nonlinear quantum device (the SHB) generates mixing between $\omega_\text{RF}$ and $\omega_\text{MW}$, as well as with all their higher harmonics. The measured signal is then dictated by the resonator, which acts as a sharp band-pass filter centred at $\omega_\text{RF}/2 \pi$, thus breaking the symmetry between the RF and MW and selecting only the component with $N=1$ and $M=0$~\cite{oakes2023multiplier,Tucker_Feldman_1985}.

While this explains the qualitatively different responses of the system to $\delta \varepsilon_\text{RF}$ and $\delta \varepsilon_\text{MW}$, it still does not provide a physical intuition for the behaviour in Fig.~\ref{fig:Lz-power-broadening}. In particular, we refer to how the QD still responds as if the MW dressing is generating \textit{copies} of the original system, with higher rungs of the Floquet ladder broadening in the fan-like response (Fig.~\ref{fig:Lz-power-broadening}e) we see at zero MW power (Fig.~\ref{fig:Lz-power-broadening}c). 
However, this is immediately understood from the fact that $H_\text{MW}$ commutes with both $H_0$ and $H_\text{RF}$. Therefore, the time propagator of the quantum dynamics $\exp\left(-\frac{\rm i}{\hbar} \int H(t')\right)$ neatly separates into a single SHB (driven by the RF probe) and the MW dressing. Materially, this translates to the fact that the eigenstates of the (non-periodic) time-dependent Hamiltonian still take the Floquet-like form 
\begin{equation}
    \begin{aligned}
    \ket{\phi(t)} =& e^{\text{i}\varepsilon_0t/\hbar} \sum_{m=-\infty}^{\infty} \sum_{n=-\infty}^{\infty} J_m\left( \frac{\delta\varepsilon_\text{MW}}{\hbar\omega_\text{MW}}\right) \\
    &J_n\left( \frac{\delta\varepsilon_\text{RF}}{\hbar\omega_\text{RF}}\right)
    e^{\text{i}m\omega_\text{MW} t} e^{\text{i}n\omega_\text{RF} t}
    \ket{\phi_{m,n}} ,
    \end{aligned}
    \label{eq:Double_Floquet_State}
\end{equation}
\noindent which resembles Eq.~(\ref{eq:Floquet_State}) but describes \textit{two} interdigitated Floquet ladders, one for the RF and one for the MW. In particular, given the large difference in photon energies in the experiment, it is most useful to consider one RF ladder originating from \textit{each} MW rung (Fig.~\ref{fig:Lz-power-broadening}a).
We can now formally consider the interaction with the reservoir as Floquet scattering, causing transitions between different rungs~\cite{Peri_2024,Dann_Levy_Kosloff_2018,Ikeda_Chinzei_Sato_2021,Mori_2023,Koski2018,Bilitewski_Cooper_2015,Kohler_1997,Platero_Aguado_2004}, where, as stated above, we are only sensitive to $\ket{\phi_{m,n}} \rightarrow \ket{\phi_{m,n\pm 1}}$. 
Since we must only consider transitions with the \textit{same} $m$ (i.e., $M=0$), we can still write the admittance of the QD as the sum in Eq.~(\ref{eq:Y_MW}), only needing to substitute the small-signal response $Y_0$ with its power-broadened form
\begin{equation}
\begin{aligned}
    Y_{PB}(\varepsilon_0) = 2\hbar\frac{(\alpha e)^2}{\delta \varepsilon_\text{RF}} \frac{\Gamma \omega_\text{RF}^2}{\omega_\text{RF} - \text{i} \Gamma}
    \sum_{n= 1}^{+ \infty} n J_n^2\left( \frac{\delta \varepsilon}{\hbar \omega}\right) \\
    \cdot \left(\mathcal{F}(\varepsilon_0 + n \hbar \omega_\text{RF}) - \mathcal{F}(\varepsilon_0- n \hbar \omega_\text{RF})\right)
\end{aligned}
\label{eq:Y_PB}
\end{equation}
\noindent 
where $\mathcal{F}(\varepsilon_0)$ is the antiderivative of $\mathcal{F}'(\varepsilon_0)$, i.e., the convolution of the Fermi-Dirac function of the reservoir and the Lorentzian density of states of the QD. 
Notably, also Eq.~(\ref{eq:Y_PB}) takes the form of a superposition of Bessel functions, similarly to the dressing caused by the MW, the difference with Eq.~(\ref{eq:Y_MW}) originating from the fact that we are selecting transitions between \textit{adjacent} RF rungs via the resonator. 

Moreover, Eq.~(\ref{eq:Y_PB}) highlights how, unlike the previous experiments, increasing the RF probe gives rise to a \textit{coherent} source of broadening. This becomes apparent in Fig.~\ref{fig:Lz-power-broadening}e, where we apply $\delta \varepsilon_\text{MW} \approx 2 \hbar \omega_\text{MW}$ to strongly suppress the $m=0$ rung in favour of the $m=\pm1$ (purple line in Fig.~\ref{fig:Lz-power-broadening}b). Increasing the RF amplitude such that $\delta \varepsilon_\text{RF} > \hbar \omega_\text{MW}$ then causes the two rungs to interfere constructively, giving rise to an enhancement of the signal, in complete agreement with our simulation.

Lastly, we note the absence of additional interference fringes in the superposition of the two power-broadening fans, which have instead been observed elsewhere in the literature~\cite{Zalba_Shevchenko_2016,Koski2018,Gu_2023}. 
This is caused by the absence of any avoided crossing between charge states, which would lead to coherent accumulation of a diabatic phase via Landau-Zener-St\"uckelberg-type processes. 
In our system, instead, the origin of the interference is caused by the different photon number states in the Floquet rungs~\cite{Koski2018,Rudner_Lindner_2020}, which can be pictured as a gap opening at the edge of the Floquet-Brillouin zone where different Floquet rungs come on to resonance~\cite{Lu_2022,Rudner_Lindner_2020, Rudner_Lindner_2020_2}. Since $\omega_\text{MW}$ is incommensurate with $\omega_\text{RF}$, the separate (blue) \text{rf} ladders originating from the (green) MW rungs in Fig.~\ref{fig:Lz-power-broadening}a remain interdigitated, avoiding further Floquet resonances and thus the opening of additional gaps.
This highlights the stochastic nature of the dispersive signal, as the reservoir incoherently exchanges holes with the QD. Yet, the quantum system still shows signs of coherent interference between the two separate Floquet ladders.

\section{Conclusion}


Our work has investigated the non-equilibrium dynamics of a dressed hole QD stochastically exchanging charges with a reservoir. We have shown the emergence of interference fringes due to coherent multiphoton transitions, revealing the underlying structure of Floquet states. We re-emphasize that the system lacks an avoided anticrossing, as is commonly observed in double-QD systems, highlighting the novelty of our observation. Overall, this work demonstrates how, despite tunnelling to and from a reservoir completely randomizing the hole charge phase, this does not preclude quantum interference phenomena, allowing instead their dispersive sensing via the system's electrical response.
Having indicated the conditions under which single-QD Floquet interferometry is observable, we predict that the technique will be widely used for fast and accurate capacitance matrix extraction of these QD-reservoir systems. In particular, the photon line spacing provides a direct measurement of the gate lever arm, which, combined with cross-capacitance measurements, reveals the full set of electrostatic parameters. The technique provides an alternative to the experimentally time-consuming methods such as thermal-broadening measurements~\cite{Ahmed2018b} needed to fully characterise the system. Future measurements could be directed to exploring the role of multi-microwave excitations and the potential role of a quantized density of states in the reservoir.

\section*{Methodology}
\textbf{Fabrication details.} The transistor used in this study consists of a single-gate silicon-on-insulator (SOI) nanowire transistor with a channel width of 120\,nm, a length of 60\,nm and height of 8\,nm. The SOI consists of a buried oxide 145\,nm thick and an 8\,nm silicon layer with a boron doping density of $5\cdot 10^{17}$\,cm$^{-3}$. The silicon layer was patterned to create the channel using optical lithography, followed by a resist-trimming process. The transistor gate stack consists of 1.9 nm HfSiON capped by 5 nm TiN and 50\,nm polycrystalline silicon, leading to a total equivalent oxide thickness of 1.3\,nm. After gate etching, a Si$_3$N$_4$ layer (10\,nm) was deposited and etched to form a first spacer on the sidewalls of the gate, then 18-nm Si raised source and drain contacts were selectively grown before source/drain extension implantation and activation annealing. A second spacer was formed, followed by source/drain implantations, an activation spike anneal and salicidation (NiPtSi). The nanowire quantum device and superconducting resonator were connected via aluminium bond wires.

\textbf{Measurement set-up.} Measurements were performed at the base temperature of a dilution refrigerator ($T\sim 10$\,mK). Low-frequency signals ($V_\text{g}$, $V_\text{bg}$) were applied through cryogenic filters, while radio-frequency and high-frequency microwave tones were applied through filtered coaxial lines to a coupling capacitor connected to the RF resonator, or through an on-PCB (printed circuit-board) bias-T connected to the Source of the transistor, respectively. The MW tones were produced by an Anritsu MG3694C Signal Generator. The resonator consists of a NbTiN superconducting spiral inductor ($L \sim 30$\,nH), coupling capacitor ($C_c \sim 40$\,fF) and low-pass filter fabricated by Star Cryoelectronics. For exact details see Ref.~\cite{vonhorstig2024}. The PCB was made from RO4003C 0.8\,mm thick with an immersion silver finish. The reflected RF signal was amplified at 4\,K and room temperature, followed by quadrature demodulation (Polyphase Microwave AD0540B), from which the amplitude and phase of the reflected signal were obtained (homodyne detection).


\section*{Acknowledgement}
 This research was supported by European Union's Horizon 2020 research and innovation programme under grant agreement no.\ 951852 (QLSI), and by the UK's Engineering and Physical Sciences Research Council (EPSRC) via the Cambridge NanoDTC (EP/L015978/1). F.E.v.H. acknowledges funding from the Gates Cambridge fellowship (Grant No. OPP1144). M.F.G.Z. acknowledges a UKRI Future Leaders Fellowship [MR/V023284/1]. L.P. acknowledges the Winton Programme for the Physics of Sustainability.

\appendix

\section{Coherent interactions of a different quantum dot with a reservoir}\label{app:sec:LZ-vs-freq-better}
\begin{figure*}
    \centering
    \includegraphics[width=1\textwidth]{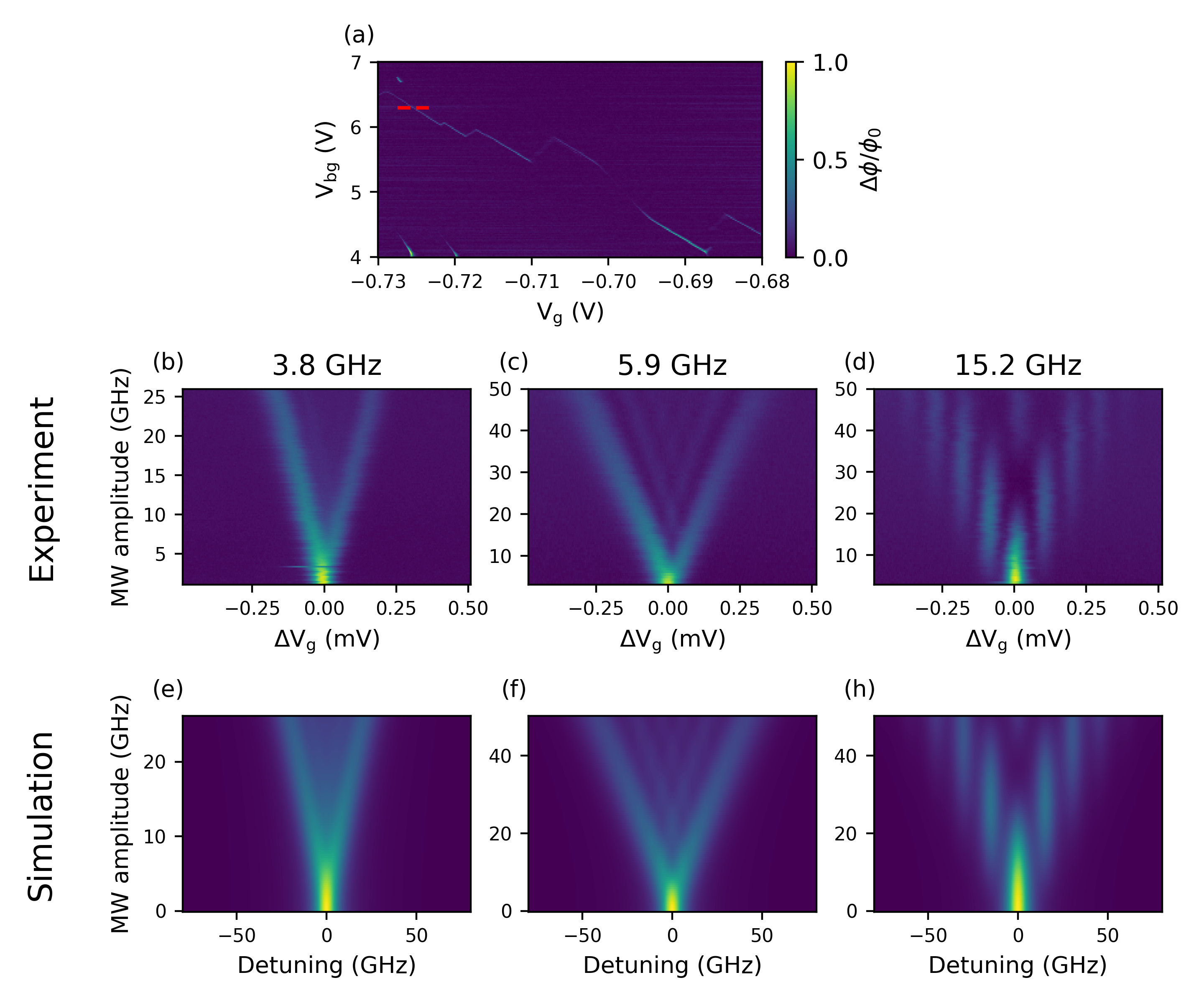}
    \caption{(a) Stability diagram showing the dot--reservoir transition. Measurements are taken at the dashed line indicated. (b-d) Experimental data of Floquet interferometry carried out at the three different frequencies indicated. (e-h) Simulated response matching the above experimental results.}
    \label{fig:app:LZ-vs-freq-better}
\end{figure*}

To demonstrate the reproducibility of our observations, we repeat the experiment as described for a different QD in the same device, this time produced by a boron dopant (see Fig.~\ref{fig:LZ-vs-freq}). In Fig.~\ref{fig:app:LZ-vs-freq-better}, we show the stability diagram of the transition showing a DRT intersected by several transitions with other QDs. We focus on the DRT indicated by the dashed line.

We now apply MW tones of three different frequencies ($\omega_\text{MW}/2\pi = 3.8$\,GHz, 5.9\,GHz and 15.2\,GHz) and measure the phase response of the resonator as a function of gate voltage and MW power (Fig.~\ref{fig:app:LZ-vs-freq-better}b-d). As the frequency is increased, we note increasingly more distinct interference patterns. We simulate these results and find the simulation to be in good agreement with the data.

\section{Coherent interactions of a quantum dot in a different device with a reservoir}\label{app:sec:LZ-vs-freq-altdevice}

\begin{figure*}
    \centering
    \includegraphics[width=1\textwidth]{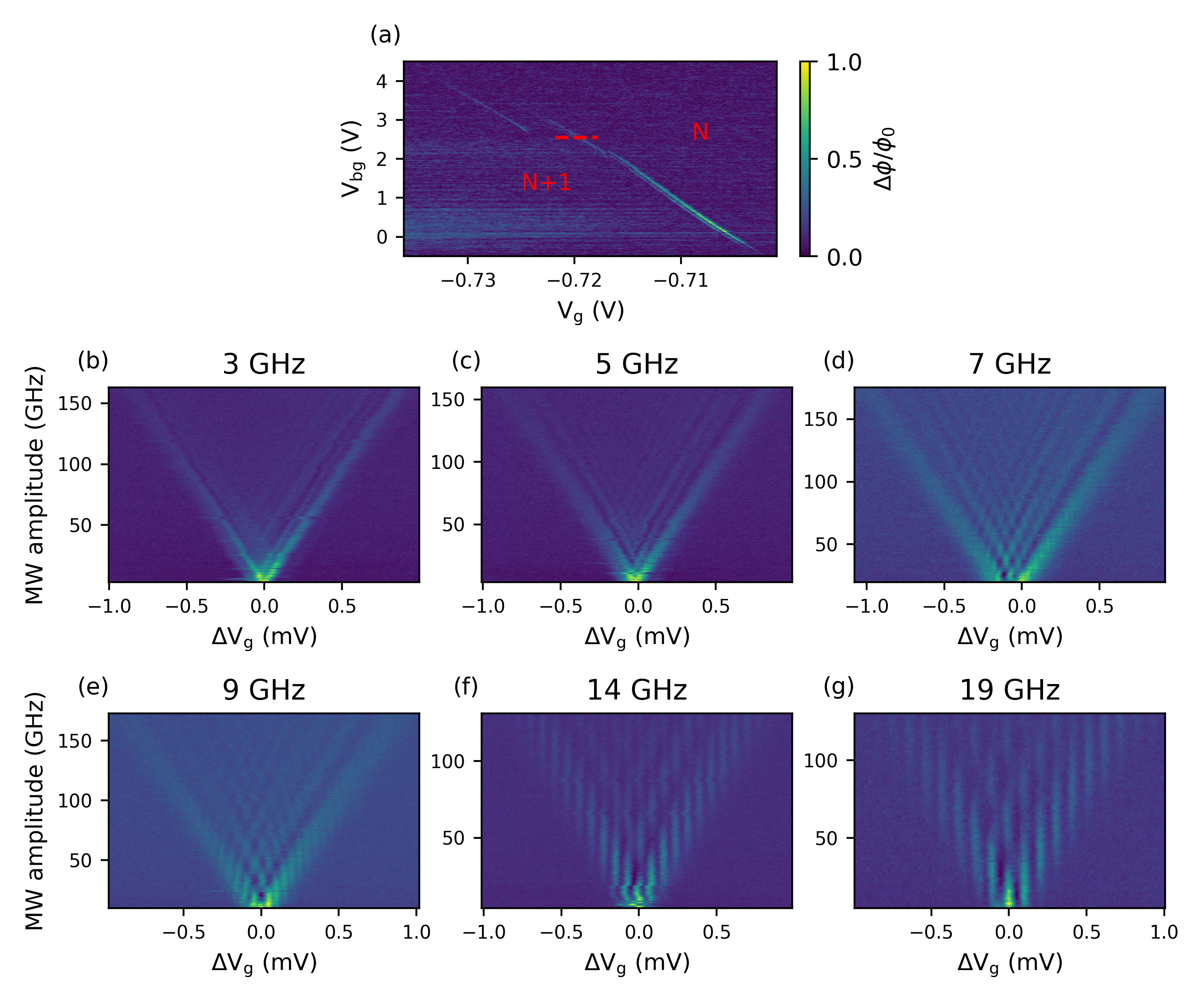}
    \caption{(a) Stability diagram showing the dot--reservoir transition and nominal charge occupation $N$ or $N+1$. Measurements are taken at the dashed line indicated. (b-g) Experimental data of Floquet interferometry carried out at the six different frequencies indicated.}
    \label{fig:app:LZ-vs-freq-altdevice}
\end{figure*}

To further illustrate the reproducible nature of our results, we repeat the experiment for a QD in a different device (see Fig.~\ref{fig:LZ-vs-freq}). In Fig.~\ref{fig:app:LZ-vs-freq-altdevice}a, we show the stability diagram of the transition, showing a DRT intersected by several transitions with other QDs. We focus on the DRT indicated by the dashed line.

We now apply MW tones of six different frequencies and measure the phase response of the resonator as a function of gate voltage and MW power (Fig.~\ref{fig:app:LZ-vs-freq-better}b-g). As the frequency is increased, we note increasingly more distinct interference patterns.

\section{Explicit Calculation of the Admittance of a Dressed Single-hole Box}
\label{app:Maths}

In this section we sketch the derivation of the high-frequency admittance of the doubly dressed single-hole box (SHB), effectively extending the formalism in Ref.~\cite{Peri_2024} to the MW-driven case.

To describe the quantum dynamics of the SHB, we can picture it as composed of four subsystems: a single level in the QD, a charge reservoir (with temperature $T_\text{h}$), and two separate photon sources, the RF and the MW. 
Each component can be described by a Hamiltonian,
\begin{align}
    &H_\text{QD} = \varepsilon_0~ d^\dagger d\\
    &H_\text{R} = \sum_\epsilon ~\epsilon ~c_\epsilon^\dagger c_\epsilon\\
    &H_\text{RF} = \hbar \omega_\text{RF} ~ a^\dagger a\\
    &H_\text{MW} = \hbar \omega_\text{MW} ~ b^\dagger b ,
\end{align}
\noindent
where $d$, $c$, $a$ and $b$ represent the destruction operators of the QD, reservoir, RF and MW, respectively. 
The interactions between them, furthermore, can be modelled as
\begin{align}
  &H_\text{QD-R} = \sum_\epsilon ~ V c_\epsilon d^\dagger + V^* c_\epsilon^\dagger d\\
  &H_\text{QD-RF} = g_\text{RF}(a + a^\dagger) d^\dagger d\\
  &H_\text{QD-MW} = g_\text{MW}(b + b^\dagger) d^\dagger d ,
\end{align}
\noindent
where we have considered the experimentally relevant case of longitudinal QD-photon coupling. 
The parameter $g_\text{RF}$($g_\text{MW}$) is the coherent QD-RF (QD-MW) coupling, and $V_\epsilon$ is the QD-reservoir coupling, which, in the wide-band approximation, we consider independent of the reservoir energy $\epsilon$~\cite{Sowa_Mol_Briggs_Gauger_2018,Sowa_Lambert_Seideman_Gauger_2020}.  The Lindblad approximation is thus justified, assuming that the thermalization time of a hole in the reservoir is much faster than the total charge tunneling rate to or from the QD. The latter reads $\Gamma = |V|^2 \mathcal{D}$, with $\mathcal{D}$ being the (flat) density of states in the reservoir.
The Hamiltonian naturally defines the (semiclassical) oscillation amplitude in terms of the time evolution of the field operator, considered in a coherent state, according to~\cite{Peri_2024,peri2023unified},
\begin{equation}
    \braket{g_\text{RF}(a + a^\dagger)}(t) = \delta \varepsilon_\text{RF} \cos{\omega_\text{RF} t}
    \label{eq:amplitude_quantum}
\end{equation}
\noindent
and similarly for the MW. We shall note that in this work we always remain in the regime of weak coupling between the QD and the radiation ($g \ll \hbar \omega$), so that the effect of spontaneous emission~\cite{Peri_2024,Mori_2023,QuantumNoise}, as well as strong-coupling phenomena~\cite{Gu_2023,Platero_Aguado_2004,Wilner_Wang_Thoss_Rabani_2015}, can be neglected.

To establish links with Floquet theory, which we have relied upon in the main text, we can perform a Lang-Frisov transformation and study the problem in the polaron frame of reference~\cite{Jang_2022,Xu_Cao_2016,Wilner_Wang_Thoss_Rabani_2015}.
This is achieved by defining the operator
\begin{equation}
    S_\text{RF} = - \frac{g_\text{RF}}{\hbar \omega_\text{RF}} \parens{a^\dagger - a} d^\dagger d
    \label{eq:polaron_operator}
\end{equation}
\noindent
and considering the canonical transformation $e^{-S_\text{RF}} H e^{S_\text{RF}}$, and similarly for the MW. 
It is important here to note that, as stressed in Section~\ref{sec:large-RF}, $[H_\text{QD-RF}, H_\text{QD-MW}] =0$, and, thus $[S_\text{QD-RF}, S_\text{QD-MW}] =0$. Therefore, the double dressing of the QD, or, more precisely, the two canonical transformations in the respective polaron frames, can be considered interchangeably and without any additional interferences as $e^{S_\text{RF}}e^{S_\text{MW}} = e^{S_\text{MW}}e^{S_\text{RF}} =  e^{S_\text{RF}+S_\text{MW}}$. 

The Lindblad master equation can now be derived according to the standard theory~\cite{Peri_2024,Yan_1998}. Notably, in Ref.~\cite{Peri_2024} we show how the dynamics of the SHB can be considered without loss of generality as a two-level system, corresponding to the QD being empty or occupied.
The unitary part of its dynamics is described by the Hamiltonian in Eq.~(\ref{eq:Hamiltonian}), while the (stochastic) interaction with the reservoir can be described by the jump operators $L_\pm = (\sigma_x \pm \text{i} \sigma_y)/2$ linked to the corresponding tunnel rates $\Gamma_{\pm}(t)$ tunneling in/out of the QD, with $\Gamma_+(t) + \Gamma_-(t) = \Gamma$ because of the conservation of charge.
In this description, the probability of occupation of the QD satisfies
\begin{equation} 
    \frac{\text{d}}{\text{d}t} P + \Gamma P = \Gamma_- (t) ,
    \label{eq:pop_ME}
\end{equation}
\noindent
from which the gate current reads $I_\text{g}(t) = \alpha e \frac{\text{d}}{\text{d}t} P(t)$ and the admittance at the resonator frequency $\omega_\text{RF}/2 \pi$~\cite{Peri_2024}
\begin{equation}
    Y = \frac{\alpha e \omega_\text{RF}}{\pi\delta \varepsilon_\text{RF}} \int_0^{\frac{2 \pi}{\omega_\text{RF}}} e^{\text{i} \omega_\text{RF}t'} I_\text{g}(t') \text{d}t'.
    \label{eq:Y_from_I}
\end{equation}

After tracing over the degrees of freedom not belonging to the QD, and the repeated application of Jacobi-Anger relation, we arrive at the results 
\begin{widetext}
    \begin{equation}
        \Gamma_-(t) = \Gamma
        \sum_{m,M,n,N=-\infty}^{\infty}
        \tilde{J}_{n,N}\left(\frac{\delta \varepsilon_\text{RF}}{\hbar \omega_\text{RF}}\right)
        \tilde{J}_{m, M}\left(\frac{\delta \varepsilon_\text{MW}}{\hbar \omega_\text{MW}}\right) \mathcal{F}(\varepsilon_0 + n \hbar \omega_\text{RF} + m \hbar \omega_\text{MW})
        \cos{(N\omega_\text{RF}t + M\omega_\text{MW}t)} ,
    \label{eq:Gamma_t}
    \end{equation}
\end{widetext}
\noindent
where we define $\tilde{J}_{n,N}(x) = J_n(x)J_{n+N}(x)$ and 
\begin{equation}
    \mathcal{F}(\varepsilon_0)= \frac{1}{e^\frac{\xi}{k_\text{B} T_\text{h}}+1} * \frac{\Gamma/\pi}{\Gamma^2 + (\xi - \varepsilon_0)^2 }
\end{equation}
\noindent
is the convolution of the Fermi-Dirac distribution of the reservoir and the Lorentzian effective density of states of the metastable QD (i.e., the antiderivative of Eq.~(\ref{eq:F_broad})). 
Notably, already in the tunnel rate, we see the quantum mixing effect discussed in Section~\ref{sec:large-RF}. Moreover, we also see how, before the band-pass filtering of the resonator, the effect of the RF probe is indistinguishable from the MW dressing, as we would expect form the general theory.
Combining Eq.~(\ref{eq:pop_ME}) and Eq.~(\ref{eq:Gamma_t}), it is then trivial to find the gate current expressed in the form of Eq.~(\ref{eq:gate_current}), where we also have derived 
\begin{equation}
\begin{aligned}
    \beta_{M,N} = \frac{\Gamma (M\omega_\text{MW} + N\omega_\text{RF})}{\sqrt{\Gamma^2 +(M\omega_\text{MW} + N\omega_\text{RF})^2 }} \sum_{m,n=-\infty}^{\infty}
    \tilde{J}_{n,N}\left(\frac{\delta \varepsilon_\text{RF}}{\hbar \omega_\text{RF}}\right)\\
    \tilde{J}_{m, M}\left(\frac{\delta \varepsilon_\text{MW}}{\hbar \omega_\text{MW}}\right) 
    \mathcal{F}(\varepsilon_0 + n \hbar \omega_\text{RF} + m \hbar \omega_\text{MW}).
\end{aligned}
\end{equation}

Equation~(\ref{eq:Y_from_I}) now mathematically takes the role of the resonator, effectively selecting only the components centered in frequency space around $\omega_\text{RF}$. 
Carrying out the Fourier transform and making use of the fact that $J_{m+1}(x) + J_{m-1}(x) = 2\frac{m}{x} J_m(x)$, we have derived Eqs.~(\ref{eq:Y_MW}) and (\ref{eq:Y_PB}). Lastly, we can exploit the fact that, for small RF amplitude, only the $n=0$ rung of the RF ladder will count in the sum in Eq.~(\ref{eq:Y_PB}) to show that 
\begin{equation}
    Y_{PB}(\varepsilon_0) \xrightarrow[\delta \varepsilon_\text{RF} \rightarrow 0]{} Y_{0}(\varepsilon_0)
\end{equation}
\noindent
as, if $\hbar \omega_\text{RF} \ll h \Gamma, k_\text{B} T_\text{h}$, then 
\begin{equation}
    \frac{\left(\mathcal{F}(\varepsilon_0 +\hbar \omega_\text{RF}) - \mathcal{F}(\varepsilon_0 - \hbar \omega_\text{RF})\right)}{2 \hbar \omega} \approx \mathcal{F}'(\varepsilon_0).
\end{equation}

As a final remark, and for ease of replicating of the simulations presented in this work, we point out how it is computationally convenient to express the convolution between the Fermi-Dirac function and the Lorentzian in terms of the polygamma functions, as 
\begin{equation}
    \mathcal{F}(\varepsilon_0) = \frac{1}{2} - \frac{1}{\pi} \Im\left[\psi_0\left(\frac{1}{2} + \text{i} \frac{\varepsilon_0}{2 \pi k_\text{B} T_\text{h}}+ \frac{h \Gamma}{2 \pi k_\text{B} T}\right)\right] ,
\end{equation} 
and
\begin{equation}
    \mathcal{F}'(\varepsilon_0) = \frac{2}{\pi^2} \Re\left[\psi_1\left(\frac{1}{2} + \text{i} \frac{\varepsilon_0}{2 \pi k_\text{B} T_\text{h}}+ \frac{h \Gamma}{2 \pi k_\text{B} T}\right)\right] ,
\end{equation} 
\noindent
where $\psi_{0(1)}$ is the digamma (trigamma) function~\cite{Peri_2024}.

\section*{Competing Interests}

The authors declare no competing financial or non-financial interests.

\section*{Data Availability}

The data that support the plots within this article and other findings of this study are available from the corresponding authors upon reasonable request.

\section*{Code Availability}

Code used for the simulations in this work can be found at \url{https://github.com/LorenzoPeri17/Beetroot}.

\bibliography{General}

\end{document}